\documentclass{pasj00}
\SetRunningHead{Tsujimoto \& Shigeyama}{History of Dwarf 
Spheroidal Galaxies}

\def\cm3{{\rm ~cm}^{-3}}
\def\ltsima{$\; \buildrel < \over \sim\;$}
\def\ltsim{\lower.5ex\hbox{\ltsima}}
\def\gtsima{$\; \buildrel > \over\sim \;$}
\def\gtsim{\lower.5ex\hbox{\gtsima}}

\author{Takuji \textsc{Tsujimoto}} 
\affil{National Astronomical Observatory, Mitaka-shi, 
Tokyo 181-8588, Japan}
\email{taku.tsujimoto@nao.ac.jp} 
\and
\author{Toshikazu \textsc{Shigeyama}}
\affil{Research Center for the Early Universe, Graduate 
School of Science, University of Tokyo, Bunkyo-ku, Tokyo 113-0033, 
Japan}
\email{shigeyama@resceu.s.u-tokyo.ac.jp}

\title{HISTORY OF MILKY WAY DWARF SPHEROIDAL GALAXIES IMPRINTED ON
ABUNDANCE PATTERNS OF NEUTRON-CAPTURE ELEMENTS}

\Received{5 April 2002}
\Accepted{24 April 2002}
\KeyWords{galaxies: abundances --- galaxies: evolution --- stars:
abundances --- galaxies: individual (Draco, Sextans, Ursa Minor) ---
supernovae: general --- supernova remnants}

\begin{document}
\maketitle

\begin{abstract}

Stellar abundance pattern of $n$-capture elements such as barium is
used as a powerful tool to infer how star formation proceeded in dwarf
spheroidal (dSph) galaxies. It is found that the abundance correlation
of barium with iron in stars belonging to dSph galaxies orbiting the
Milky Way, i.e., Draco, Sextans, and Ursa Minor have a feature similar
to the barium-iron correlation in Galactic metal-poor stars. The
common feature of these two correlations can be realized by our
inhomogeneous chemical evolution model based on the supernova-driven
star formation scenario if dSph stars formed from gas with a velocity
dispersion of $\sim$ 26 km ${\rm s}^{-1}$. This velocity dispersion
together with the stellar luminosities strongly suggest that dark
matter dominated dSph galaxies.  The tidal force of the Milky Way
links this velocity dispersion with the currently observed value
$\ltsim10$ km s$^{-1}$ by stripping the dark matter in dSph
galaxies. As a result, the total mass of each dSph galaxy is found to
have been originally $\sim 25$ times larger than at present. Our
inhomogeneous chemical evolution model succeeds in reproducing the
stellar [Fe/H] distribution function observed in Sextans. In this
model, supernovae immediately after the end of the star formation
epoch can expel the remaining gas over the gravitational potential of
the dSph galaxy.

\end{abstract}

\section{INTRODUCTION}

There is no room for doubt that the recent progress on abundance
determination for numerous solar neighborhood stars promotes a better
understanding of the nature of the Milky Way because stellar abundance
patterns bring us valuable information on how the Milky Way formed and
has evolved \citep{Wheeler_89,McWilliam_97}. We have now reached the
stage where we can learn the history of other nearby galaxies from
detailed elemental abundances of individual stars.

Recently \citet[hereafter SCS01]{Shetrone_01} revealed the abundance
patterns for red giant stars belonging to three Galactic dwarf
spheroidal (dSph) galaxies, i.e., Draco, Sextans, and Ursa
Minor. These three galaxies share similar features and reside in a
similar environment (e.g., Mateo 1998). Some important
properties are shown in Table 1. Remarkable features in the abundance
patterns of these dSph stars are found in neutron-capture elements
such as barium (Ba) and europium (Eu) (SCS01): first, the observed
Ba/Eu ratios are very close to the pure $r$-process ratio; second,
there exist stars having very high $r$-process element abundances with
respect to iron (Fe). These features are shared with extremely
metal-poor halo stars in the Milky Way
\citep{McWilliam_95,McWilliam_98}.

\begin{table*}
\begin{center}
\caption{Observed Properties of Three Galactic Dwarf Spheroidal Galaxies}
\begin{tabular}{lllllll}
\hline
& \hspace{0.1cm} $<$[Fe/H]$>^a$ & \hspace{0.4cm}$\sigma_*$$^b$ 
& \hspace{0.7cm}
$L_V$$^c$ & \hspace{-0.2cm}Distance$^d$ & \hspace{-0.1cm}Core Radius $r_c$ 
& \hspace{-0.1cm}Tidal Radius $R_t$ \\ 
Galaxy & \hspace{0.7cm}dex & (km/s) & \hspace{0.7cm}($L_\odot$) 
& (kpc) & \hspace{0.2cm}(arcmin) & \hspace{0.2cm}(arcmin) \\
\hline
Draco & $-2.0\pm0.15$ & 9.5$\pm$1.5 & $(1.8\pm0.8)\times10^5$ &
$71\pm7^e$ & \hspace{0.04cm} $8'.7\pm0'.3^e$ & $49'.4\pm1'.4^e$ \\
Ursa Minor & $-2.2\pm0.1$ & 9.3$\pm1.6$ &
$(2.0\pm0.9)\times10^5$
& $69\pm4^f$ & $15'.8\pm1'.2$ & $50'.6\pm3'.6$ \\
Sextans & $-2.1\pm0.04^g$ & 6.6$\pm$0.7 &
$(4.1\pm1.9)\times10^5$
& $86\pm4$ & $16'.6\pm1'.2$ & \hspace{0.000001cm} $160'\pm50'$ \\
\hline
\end{tabular}
\end{center}

\small{Note.-- All data for which references are not indicated are
from \citet{Mateo_98}. $^a$Mean iron stellar abundance. $^b$Stellar
central velocity dispersion.  $^c$Visual Luminosity, from
\citet{Irwin_95}. $^d$Distance from the center of the Milky Way.
$^e$From \citet{Suntzeff_93}. $^f$From \citet{Odenkirchen_01}.
$^g$From \citet{Eskridge_01}.
}
\end{table*}

Needless to say that the first feature suggests that Ba in these dSph
stars is of $r$-process origin. In addition, the Ba/Eu ratio is
expected to work as a {\it cosmic clock} which tells us the stellar
age -- that is -- the timescale for the formation of dSph galaxies. In
general, the Ba/Eu ratios in stars increases with time because the
$s$-process synthesizes Ba but little Eu and it takes a much longer
time for the $s$-process to contribute to the chemical enrichment than
the $r$-process (e.g., Pagel 1997). It comes down to that the pure
$r$-process Ba/Eu ratio in stars indicates that these stars had been
formed before the $s$-process began to produce Ba. As a result, the
timescale of the star formation in these dSph galaxies needs to be
less than a few $10^8$ years. This short timescale strongly suggests
that there must be no contribution to the chemical evolution from type
Ia supernovae (SNe Ia). Thus the small Mg/Fe ratios observed in dSph
stars might be ascribed to the rapid p-capture process that converts
Mg into aluminum (Al) with the help of deep mixing in stars
\citep{Sweigart_79, Fujimoto_99}, though the abundances of Al in dSph
stars were too poorly estimated to tell if this process occurs in dSph
stars (SCS01).  To avoid the influence by nuclear reactions that might
occur inside old stars, we will extract information on history of dSph
galaxies from the abundances of elements heavier than Fe.

\begin{table*}
\begin{center}
\caption{Comparison of Elemental
Abundances between CS22892-052 and Dwarf Spheroidal Stars} 
\begin{tabular}{ccccccccc}
\hline
Star & Galaxy & [Fe/H] & [Ba/Fe] & [Ce/Fe] & [Nd/Fe] &
[Sm/Fe] & [Eu/Fe] & [Ba/Eu] \\ \hline
CS22892-052 & the Milky Way & $-$3.10 & 0.91 & 1.05 &
1.22 & 1.56 & 1.66 & $-$0.75 \\ 199 & Ursa Minor & $-$1.45 &
0.77 & 1.03 & 1.13 & 1.75 & 1.49 & $-$0.72 \\ S35 &
Sextans & $-$1.93 & 0.70 & 0.89 & 0.79 & 1.58 & -- & -- \\
\hline
\end{tabular}
\end{center}

\small{Note.-- The star K in Ursa Minor is much more enhanced
in [X/Fe] than above dSph stars. However the Ba/Eu ratio in this star is
far deviated from the value anticipated from the pure $r$-process
origin.}
\end{table*}

The second feature is very intriguing. Elements such as Ce, Nd, and Sm
as well as Eu and Ba are enhanced with respect to Fe in some dSph
stars, as shown in Figures 3 and 4 in SCS01, which reminds us of
notable neutron-capture-rich giant stars -- CS22892-052 and
CS31082-001 -- \citep{Sneden_00, Hill_01} in the Galactic halo. Table
2 clarifies the similarity in the [$r$-process/Fe] ratios of
CS22892-052 and two dSph stars. Furthermore, the correlations of
[Ba/Fe] ratios with [Fe/H] in these two types of stars are found to be
quite similar if the metallicities [Fe/H] of dSph stars are shifted by
the amount $\Delta$[Fe/H] = $-$0.6 dex. Figure 1 shows the correlation
of [Ba/Fe] with [Fe/H] for dSph stars (SCS01), together with that for
Galactic metal-poor stars ({\it crosses}; McWilliam 1998). Filled
circles represent dSph stars with [Fe/H] shifted by $\Delta$[Fe/H] =
$-$0.6 dex. An excellent coincidence between two distributions
appears: both metal-poor stars and dSph stars populate two separate
branches in the [Ba/Fe]-[Fe/H] plane, i.e., the first branch in which
stars have positive [Ba/Fe] ratios (i.e., greater than the solar
ratio), and the second branch with the upper bound at [Ba/Fe]=$-1$.
Though such a coincidence can be also obtained with
--0.8 dex$\leq$$\Delta$[Fe/H]$\leq$--0.6 dex, we prefer
$\Delta$[Fe/H]=$-0.6$ dex that will be justified in \S 4.

\citet{Tsujimoto_00} associate this feature, a bifurcation of the
observed element ratios [Ba/Fe], with two distinct SN classes: one
that synthesizes and ejects $r$-process elements and the other that
does not \citep{Tsujimoto2_00, Tsujimoto_01}, based on their
inhomogeneous chemical evolution model in which stars are born from
the matter swept up by individual supernova remnants (SNRs). This
assumption is essential to understand the observed elemental abundance
pattern in halo field stars and the chemical evolution at the early
epochs of the Milky Way \citep[Tsujimoto, Shigeyama, \& Yoshii 1999,
2000, 2002]{Shigeyama_98, Tsujimoto_98}. The common feature in the
abundance patterns of Galactic metal-poor stars and dSph stars
strongly suggests that this mechanism also works in dSph galaxies.

DSph galaxies have little gas, extremely low surface brightness, and low
metallicities. These features have been ascribed to gas removal by SN
explosions from a galaxy embedded in dark halo \citep{Dekel_86}. Their
scenario apparently reproduces some observed relations for dSph
galaxies. However the tidal force from the Milky Way must affect the
dynamics of these galaxies, which \citet{Dekel_86} ignored. Such
environmental effects are indeed imprinted on the elemental abundance
patterns observed for dSph stars (see the next section).

In this {\it letter}, we describe the star formation history of a dSph
galaxy in terms of an inhomogeneous chemical evolution model developed
for the halo of the Milky Way. The observed abundance trends seen in
neutron-capture elements and stellar abundance distribution function
(ADF) constrain our model to shed light on the history of dSph
galaxies and the environmental effects on these galaxies.

\begin{figure}
\begin{center}
\FigureFile(239bp,154bp){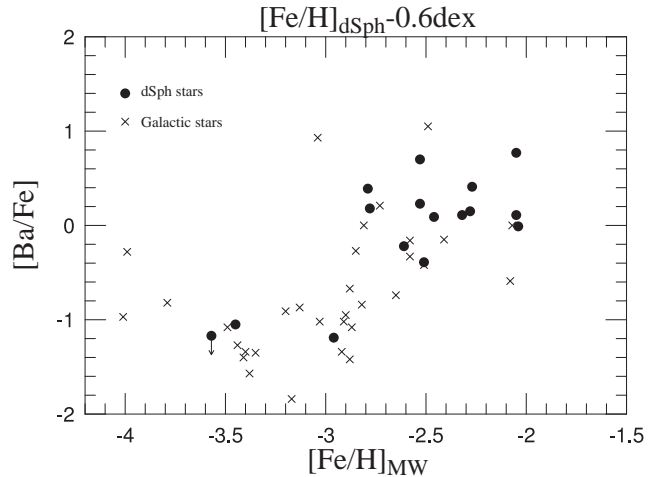}
\caption{Correlations of [Ba/Fe] with [Fe/H] for dSph stars
({\it filled circles}; SCS01) and Galactic halo field stars 
({\it crosses}; McWilliam 1998). The data of dSph stars are shifted
by the amount of $\Delta$[Fe/H]=--0.6 dex (see the text). The K star
of Ursa Minor is excluded from the plots because it is recognized to
be a carbon star, most probably self-enriched by $s$- process
material.}
\end{center}
\end{figure}

\section{HISTORY OF GALAXIES IMPRINTED ON STARS}

The metallicities of metal-poor stars are determined by two factors:
how much heavy-element mass an SN supplies and how much interstellar
matter (ISM) mass was eventually swept by a single SN explosion
\citep{Shigeyama_98}. The former is determined exclusively by each
SN. On the other hand, the latter quantity is influenced by the
environment such as the velocity dispersion $\sigma_{\rm v}$ and
density $n$ of the ISM. If SNe in the Milky Way and those in dSph
galaxies are similar, different environments should give rise to the
$\Delta$ [Fe/H] discussed above. Since the mass $M_{\rm sw}$ swept up
by an SN is much more sensitive to the velocity dispersion than to the
density; $M_{\rm sw}\propto\sigma_{\rm v}^{-9/7} n^{-0.062}$
\citep{Shigeyama_98}, it is likely that a larger velocity dispersion
in dSph galaxies at the star formation epoch enhanced the stellar
[Fe/H]. Quantitatively, $\Delta$ [Fe/H]=--0.6 dex corresponds to a
velocity dispersion of $\sigma_{\rm v}\sim$ 26 km s$^{-1}$. The
velocity dispersion $\sigma_\star$ of stars at the star formation
epoch must have a similar value in equilibrium configuration.
The present value of $\sigma_\star$ measured in Galactic dSph galaxies are
about 9 km s$^{-1}$ (see Table 1). These two significantly different
velocity dispersions can be connected in the context of a galaxy
losing mass by the tidal force of the Milky Way. The tidal radius
$R_{\rm t}$ of a dSph galaxy with a mass of $M_{\rm dSph}$ at a
distance of $D$ from the center of the Milky Way can be expressed as,
\begin{eqnarray}
R_{\rm t} &\sim& D\left({M_{\rm dSph}\over 2M_{\rm MW}}\right)^{1\over
3},\nonumber \\
&\sim& 1\, {\rm kpc}\left({M_{\rm
dSph}\over2\times10^7\,M_\odot}\right)^{1\over3}\left({D\over 70\,{\rm
kpc}}\right).
\end{eqnarray}
Here the mass of the Milky Way is denoted by $M_{\rm MW}$ and assumed
to be $3\times 10^{12}\,M_\odot$ \citep{Klessen_98}. Since the actual
sizes of dSph galaxies are also $\sim$1 kpc, dSph galaxies are likely
to be affected by the tidal force of the Milky Way. If a dSph galaxy
extends to the tidal radius and in virial equilibrium, the velocity
dispersion is given by
\begin{eqnarray}
\label{eqn:vd}
\sigma_\star &\sim& \sqrt{GM_{\rm dSph}\over R_{\rm t}}=\sqrt{GM_{\rm
MW}\over D}\left({\sqrt{2}M_{\rm dSph}\over M_{\rm MW}}\right)^{1\over
3},
\nonumber \\
&\sim&9\,{\rm km\, s}^{-1}\left({M_{\rm
dSph}\over2\times10^7\,M_\odot}\right)^{1\over3}\left({D\over 70\,{\rm
kpc}}\right)^{-{1\over 2}}.
\end{eqnarray}
The above formulae well describe results of numerical $N$-body
calculations for Galactic dSph galaxies \citep{Oh_95}. A larger
velocity dispersion at the star formation epoch inferred from the
elemental abundance pattern suggests that the total mass of the dSph
galaxy at that time must be $\sim$25 times larger than at present (see
Eq. (\ref{eqn:vd})). The timescale $|dt/d\ln M_{\rm dSph}|$ for
stripping is expressed and evaluated as
$
|dt\big/d\ln M_{\rm dSph}|={2\sqrt{3}R_{\rm
t}\big/\sigma_\star}=\sqrt{{12D^3\big/ \left(GM_{\rm MW}\right)}}
\sim 1.5-3 \,{\rm Gyr},
$
depending on the estimated mass of the Milky Way. From this timescale
and the initial mass of the dSph galaxy, the corresponding age of
Galactic dSph galaxies is estimated to range from 5 to 10 Gyr.

From an elemental abundance pattern which has been converted to
information on the velocity dispersion of gas, we have estimated the
initial dynamical (total) mass of dSph galaxies. In the same manner,
i.e., from the chemical properties imprinted on stars, we deduce the
baryonic mass initially residing in dSph galaxies. From the
conservation of Fe, we have the relation $Z_sM_s+Z_gM_g=yM_s$, where
$M_s$: the mass of stars, $Z_s$: the mean metallicity of stars, $M_g
(Z_g)$: the mass (metallicity) of the gas at the end of star
formation, and $y$: the Fe yield (see Tinsley 1980). If we define
$\varepsilon_{\rm s}$ as the mass fraction of the initial gas that has
been converted to stars in the end, then $\varepsilon_{\rm s}$ can be
expressed as $\varepsilon_{\rm s} = Z_g/(y+Z_g-Z_s)$. From the ADF for
Sextans observed by \citet{Suntzeff_93} (see Fig.~2), we have
[Fe/H]=$-2.0$ for $Z_s$, and [Fe/H]=$-1.6$ (corresponding to the
metallicity of the most metal-rich star) for $Z_g$. Adopting
$y$=0.38$Z_{{\rm Fe},\odot}$ for the Fe yield from SNe II
\citep{Tsujimoto_97}, we deduce $\varepsilon_{\rm s} \sim 0.05$. Our
numerical calculation presented in the next section confirms this
value. Thus the initial gas (baryonic) mass was 20 times larger than
the present stellar mass. Interestingly the ratio of the initial to
the present baryonic mass is close to that for dark matter. In other
words, the baryonic fraction $f_{\rm b}$ in a dSph galaxy is quite
similar at these two phases.

We try to make a rough estimate of $f_{\rm b}$ for Draco. Taking
$L_{\rm V}=1.8\times 10^5 L_{\rm V,\odot}$ \citep{Irwin_95} and
the mass-to-light ratio for old stellar populations $M_\star/L \sim 8$
\citep{Marel_91}, we obtain the present baryonic mass $\sim
1.4\times 10^6 M_\odot$. If we adopt the present dynamical mass of
$2.2\times 10^7 M_\odot$ \citep{Odenkirchen_01}, we obtain
$f_{\rm b} \sim 0.06$. Our scenario predicts that $f_{\rm b}$ was
also $\sim 0.06$ at the initial formation phase.

In our inhomogeneous chemical evolution model, the star formation is
terminated when too many SNe explode to sweep a sufficient amount of
the ISM. The number of these SNe is found to be a few hundred from our
model. These last few hundred SN explosions were able to disperse the
remaining gas over the potential well made by dark matter (see
discussion in \S\ref{sect:sfh}).

\section{INHOMOGENEOUS CHEMICAL EVOLUTION}

In this section we discuss Ba and Fe in dSph galaxies in terms of the
inhomogeneous chemical evolution model presented in
\citet{Tsujimoto2_99}. The model is based on a scenario where the
chemical evolution proceeds through a repetition of a sequence of SN
explosion, shell formation, and star formation in the matter
individual SNRs sweep up. As discussed in the previous section, the
velocity dispersion of gas is assumed to be a constant value 26 km
s$^{-1}$ because the timescale of tidal interaction is $\sim$10 times
longer than that of star formation. The free parameters in our model
are the mass fraction $x_{\rm III}$ of metal-free stars initially
formed, and the mass fraction $\epsilon$ of stars formed in the dense
shells swept up by each SNR. We simply take $x_{\rm
III}=2.5\times10^{-4}$ inferred from the Ba abundances of extremely
metal-poor stars in the Milky Way \citep{Tsujimoto2_00}, because the
results are rather insensitive to this value. Stars are assumed to be
formed following the Salpeter initial mass function.

The value of $\epsilon$ is determined in order to reproduce the
observed ADF \citep{Tsujimoto2_99}. The location of the ADF peak is
sensitive to what fraction of the gas has been converted into stars in
the end. The observed peak for Sextans at [Fe/H] $\sim -2$
\citep{Suntzeff_93} requires less efficient star formation than in the
Galactic halo the ADF of which has a peak at [Fe/H] $\sim -1.6$. A
conversion of $\sim$5 \% of the gas into stars is found to give the
best fit to the observed ADF, which is realized by
$\epsilon=1.5\times10^{-2}$.  This $\epsilon$ produces slightly more
than one star that can explode as a SN per SNR shell on
average. Figure 2a shows the ADF obtained by our model, as compared
with the data acquired by \citet{Suntzeff_93}. The result has been
convolved with a Gaussian with the dispersion of $\sigma$=0.1 dex
which is identical to the measurement error in [Fe/H]. A good
agreement with the data is obtained, though the number of the observed
stars is too small to strictly constrain models.

\begin{figure}
\begin{center}
\FigureFile(262bp,274bp){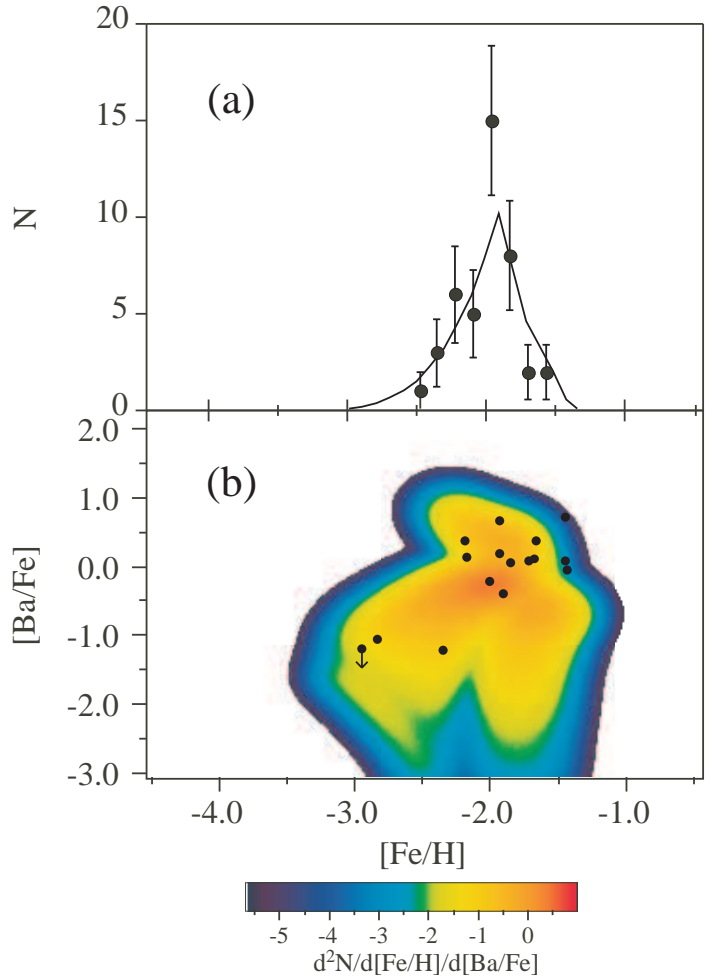}
\caption{($a$) The frequency distribution of dSph stars
against the iron abundance, compared with the observation for Sextans
\citep{Suntzeff_93}. The error bars take into account Poisson noise. 
($b$) The color-coded frequency distribution of
dSph stars in the [Ba/Fe]-[Fe/H] plane convolved with a Gaussian
having $\sigma=0.2$ dex for [Ba/Fe] and $\sigma=0.1$ dex for
[Fe/H]. Filled circles show the data of SCS01.}
\end{center}
\end{figure}

Assuming that SNe II with $M_{\rm ms}=20-25M_\odot$ are the dominant
site for $r$-process nucleosynthesis \citep{Tsujimoto2_00,
Tsujimoto_01}, we calculate the Ba evolution in dSph galaxies. Figure
2b is a color-coded predicted frequency distribution of stars in the
[Ba/Fe]-[Fe/H] plane, normalized to unity when integrated over the
entire area. In order to make a direct comparison with the data, the
frequency distribution has been convolved with a Gaussian with
$\sigma$= 0.2 dex for [Ba/Fe] and $\sigma$=0.1 dex for [Fe/H]. Our
model predicts a bent arrow-like frequency distribution of stars,
which covers the present observational data points.

\section{GAS REMOVAL}
\label{sect:sfh} In our model, the formation of new stars is terminated
when SNRs sweep up too little gas without being affected by other SNe
to form cool shells. After the end of star formation, the remaining
gas is therefore heated up by SN explosions without experiencing
effective radiative cooling. The total energy added by these last SNe
amounts to $\sim5\times10^{53}$ erg. As compared with the
gravitational binding energy of the remaining gas $E_{\rm g}\sim
GM_{\rm dSph}M_{\rm gas}/R_{\rm t}\sim 4\times10^{53}$ erg, all the
gas could be expelled by these SNe. If the energy supplied by the last
SNe were significantly smaller than $E_{\rm g}$, the gas would still
remain in the gravitational potential. Then the star formation would
recommence in the radiative cooling timescale. Such a situation would
be realized if the metallicity difference between dSph stars and
Galactic field stars was $\Delta$[Fe/H]$\approx -0.7 - -0.8$dex. In
this case, a larger velocity dispersion of gas in the proto-dSph
galaxies would result in $E_{\rm g}$\gtsim$10^{54}$ erg. No sign of
the second episode of star formation in neither the observed ADF nor
Ba/Eu ratios suggests that this is not the case.

The total number of SNe is $\sim$3,000 throughout the star formation
episode in our model.  If all the explosion energy of these SNe were
converted into kinetic energy of the gas, the mean velocity would
become 100 km s$^{-1}$ at most. Thus it is likely that our model dSph
galaxy never satisfies the criterion for SN-driven winds proposed by
\citet{Dekel_86} that SNe increase the velocity dispersion of the gas
to greater than $\sim$100 km s$^{-1}$ to escape from a dSph
galaxy. Therefore more SNe are needed to expel the gas by SN-driven
winds according to \citet{Dekel_86}. In other words, a galaxy that
satisfied this criterion would produce too much heavy-element mass to
reconcile with the mean stellar Fe abundances in Table 1.

\section{CONCLUSIONS AND DISCUSSIONS}

We show that stellar elemental abundance patterns for $n$-capture
elements have shed light on the histories of galaxies. The Ba/Eu
ratios in dSph stars severely constrain the timescale of the star
formation to be of the order of 10$^8$ yr.  Recently observed
correlation of [Ba/Fe] with [Fe/H] for dSph stars can be reconciled
with that seen in Galactic halo stars by introducing a larger velocity
dispersion in dSph galaxies at the star formation epoch, if stars are
born from individual SNRs.  These considerations lead us to our
finding that dSph stars were born from the gas with its velocity
dispersion of $\sim 26$ km s$^{-1}$, which results in a mass of the
proto-dSph galaxies $\sim$25 times larger than at present. The path to
the present dSph galaxies is likely to be controlled by the tidal
interaction with the Milky Way.

The history of Galactic dSph galaxies that we have proposed is as
follows. The proto-dSph galaxy consisted of gas and dark matter, and
was approximately 25 times as massive as a present dSph galaxy. Star
formation lasted for about $2\times10^8$yr; meanwhile,
$\sim$ 5 percent of the gas converted into stars. Afterward, the
remaining gas was dispelled by the last SN explosions. On the other
hand, dark matter which surrounds stars has lost more than 90\% of its
mass through the tidal interaction with the Milky Way.

Lighter elements might include information that has not been discussed
in this paper. Deficiencies in Ca/Fe and Ti/Fe ratios are also seen in
dSph stars (SCS01) as compared to the halo stars, though not so
pronounced as Mg/Fe. In the framework of our inhomogeneous chemical
evolution model, these deficiencies should be attributed to SNe II in
dSph galaxies that yield a relatively small amount of $\alpha$
elements as compared with Fe and/or SNe that yield a large amount of
Fe. The former SNe will not alter our scenario for dSph galaxy
evolution. In the latter case, the SNe yields will contribute to a
part of the discrepancy in Fe/H ratios between the Milky Way and dSph
galaxies and reduce the initial velocity dispersion (and mass) of
matter accordingly.

Further stellar abundance measurements not only for dSph galaxies
discussed here but also for other nearby galaxies promise an
understanding of their star formation histories more precisely than
deciphering the color-magnitude diagram of evolved stars which has
been employed so far.

\bigskip
We are grateful to the anonymous referee for making useful comments.
This work has been partly supported by a Grant-in-Aid for Scientific
Research (11640229, 12640242) of the Ministry of Education, Culture,
Sports, Science, and Technology in Japan.


\begin{thebibliography}{}
\bibitem[Dekel \& Silk(1986)]{Dekel_86}
Dekel, A. \& Silk, J. 1986, \apj, 303, 39
\bibitem[Eskridge \& Schweitzer(2001)]{Eskridge_01}
Eskridge, P. B. \& Schweitzer, A. E. 2001, AJ, 122, 3106
\bibitem[Fujimoto, Aikawa, \& Kato(1999)]{Fujimoto_99}
Fujimoto, M. Y., Aikawa, M., \& Kato, K. 1999, \apj, 519, 733
\bibitem[Hill et al.(2001)]{Hill_01}
Hill, V., Plez, B., Cayrel, R., \& Beers, T. C. 2001, in Astrophysical
Ages and Time Scales, eds. T. Von Hippel, C. Simpson, \&  N. Manset
(ASP: San Francisco), 316
\bibitem[Irwin \& Hatzdimitriou (1995)]{Irwin_95}
Irwin, M., \& Hatzdimitriou, D. 1995, MNRAS, 277, 1354
\bibitem[Klessen \& Kroupa(1998)]{Klessen_98}
Klessen, R. S., \& Kroupa, P. 1998, \apj, 498, 143
\bibitem[Mateo(1998)]{Mateo_98}
Mateo, M. L. 1998, \araa, 36, 435
\bibitem[McWilliam et al.(1995)]{McWilliam_95}
McWilliam, A., Preston, G. W., Sneden, C., \& Searle, L. 1995, \aj, 109,
2757
\bibitem[McWilliam(1997)]{McWilliam_97}
McWilliam, A. 1997, ARA\&A, 35, 503
\bibitem[McWilliam(1998)]{McWilliam_98}
McWilliam, A. 1998, \aj, 115, 1640
\bibitem[Odenkirchen et al.(2001)]{Odenkirchen_01}
Odenkirchen, M., et al. 2001, \aj, 122, 2538
\bibitem[Oh, Lin, \& Aarseth(1995)]{Oh_95}
Oh, K. S., Lin, D. N. C., \& Aarseth, S. J. 1995, \apj, 442, 142
\bibitem[Pagel(1997)]{Pagel_97}
Pagel, B. E. J. 1997, Nucleosynthesis and Chemical Evolution of Galaxies
(Cambridge: Cambridge Univ. Press)
\bibitem[Shetrone, C\^{o}t\'{e}, \& Sargent(2001)]{Shetrone_01}
Shetrone, M. D., C\^{o}t\'{e}, P., \& Sargent, W. L. 2001, \apj, 548, 592
(SCS01)
\bibitem[Shigeyama \& Tsujimoto(1998)]{Shigeyama_98}
Shigeyama, T., \& Tsujimoto, T. 1998, \apj, 507, L135
\bibitem[Sneden et al.(2000)]{Sneden_00}
Sneden, C., Cowan, J. J., Ivans, I. I., Fuller, G. M., Burles, S.,
Beers, T. C., \& Lawler, J. E. 2000, \apj, 533, L139
\bibitem[Suntzeff et al.(1993)]{Suntzeff_93}
Suntzeff, N. B., Mateo, M., Terndrup, D. M., Olszewski, E. W.,
Geisler, D., \& Weller, W. 1993, ApJ, 418, 208
\bibitem[Sweigart \& Mengel(1979)]{Sweigart_79}
Sweigart, A. V., \& Mengel, J. G. 1979, \apj, 229, 624
\bibitem[Tinsley (1980)]{Tinsley_80}
Tinsley, B. M. 1980, Fundam. Cosmic Phys., 5, 287
\bibitem[Tsujimoto et al.(1997)]{Tsujimoto_97}
Tsujimoto, T., Yoshii, Y., Nomoto, K., Matteucci, F., Thielemann,
F.-K., \& Hashimoto, M. 1997, ApJ, 483, 228
\bibitem[Tsujimoto \& Shigeyama(1998)]{Tsujimoto_98}
Tsujimoto, T., \& Shigeyama, T. 1998, \apj, 508, L151
\bibitem[Tsujimoto, Shigeyama, \& Yoshii(1999)]{Tsujimoto_99}
Tsujimoto, T., Shigeyama, T., \& Yoshii, Y. 1999, \apj, 519, L63
\bibitem[Tsujimoto, Shigeyama, \& Yoshii(2000)]{Tsujimoto_00}
Tsujimoto, T., Shigeyama, T., \& Yoshii, Y. 2000, \apj, 531, L3
\bibitem[Tsujimoto \& Shigeyama(2001)]{Tsujimoto_01}
Tsujimoto, T., \& Shigeyama, T. 2001, \apj, 561, L97
\bibitem[Tsujimoto, Shigeyama, \& Yoshii(2002)]{Tsujimoto_02}
Tsujimoto, T., Shigeyama, T., \& Yoshii, Y. 2002, \apj, 565, 1011
\bibitem[van der Marel (1991)]{Marel_91}
van der Marel, R. P. 1991, MNRAS, 253, 710
\bibitem[Wheeler, Sneden, \& Truran(1989)]{Wheeler_89}
Wheeler, J. C., Sneden, C., \& Truran, J. W. 1989, ARA\&A, 27, 279
\bibitem[Tsujimoto et al.(1999)]{Tsujimoto2_99}
\bibitem[Tsujimoto et al.(2000)]{Tsujimoto2_00}
\bibitem[Oh et al.(1995)]{Oh2_95}
\end{thebibliography}
\end{document}